# ARTICLE

Check for updates

# Multilayer hazes over Saturn's hexagon from Cassini ISS limb images

A. Sánchez-Lavega  [1✉], A. García-Muñoz[2], T. del Río-Gaztelurrutia  [1], S. Pérez-Hoyos  [1], J. F. Sanz-Requena  [3], R. Hueso  [1], S. Guerlet[4] & J. Peralta  [5]

In June 2015, Cassini high-resolution images of Saturn's limb southwards of the planet's hexagonal wave revealed a system of at least six stacked haze layers above the upper cloud deck. Here, we characterize those haze layers and discuss their nature. Vertical thickness of layers ranged from 7 to 18 km, and they extended in altitude ~130 km, from pressure level 0.5 bar to 0.01 bar. Above them, a thin but extended aerosol layer reached altitude ~340 km (0.4 mbar). Radiative transfer modeling of spectral reflectivity shows that haze properties are consistent with particles of diameter 0.07–1.4 μm and number density 100–500 cm$^{-3}$. The nature of the hazes is compatible with their formation by condensation of hydrocarbon ices, including acetylene and benzene at higher altitudes. Their vertical distribution could be due to upward propagating gravity waves generated by dynamical forcing by the hexagon and its associated eastward jet.

[1] Departamento Física Aplicada I, Escuela de Ingeniería de Bilbao, Universidad del País Vasco UPV/EHU, Bilbao, Spain. [2] Zentrum für Astronomie und Astrophysik, Technische Universität Berlin, Berlin, Germany. [3] Departamento de Física Teórica, Atómica y Optica, Universidad de Valladolid, Valladolid, Spain. [4] Laboratoire de Meteorologie Dynamique/Institut Pierre-Simon Laplace (LMD/IPSL), Sorbonne Universite, Centre National de la Recherche Scientifique (CNRS), Ecole Polytechnique, Ecole Normale Superieure (ENS), Paris, France. [5] Institute of Space and Astronautical Science, Japan Aerospace Exploration Agency, Kanagawa, Japan. ✉email: agustin.sanchez@ehu.eus

Saturn's hexagon is a unique and persistent wave. It was first observed at the upper cloud level by the Voyager 1 and 2 during their flybys of the planet in 1981[1]. Since 1991, it has been imaged using the Hubble Space Telescope (HST) and ground-based telescopes and, between 2006 and 2017, with the Cassini Imaging Science Subsystem (ISS) and Visual and Infrared Mapping Spectrometer (VIMS)[2–9]. The hexagon stands out at visible and adjacent wavelengths (~0.2–1 microns) by reflecting sunlight at the upper clouds (pressure level ~0.5–1 bar), and by transmission and opacity of the clouds at ~5 microns (pressure level ~2–3 bar)[10]. Within the hexagonal wave, winds blow in an intense and narrow jet with maximum speeds of 120 ms$^{-1}$[1,5–8]. The hexagon also manifests itself in the temperature field in the upper troposphere at ~0.1 bar and in the stratosphere (5–0.5 mbar), more than ~300 km above the clouds[11]. To date, its recorded lifetime is >38 years, more than one Saturn's year (29.4571 Earth years), showing that it has survived the strong seasonal insolation cycle at the top of the atmosphere[9–13].

In June 2015, the Cassini spacecraft obtained images of the planet limb at a high spatial resolution of 1–2 km/pixel, covering a broad wavelength range from 225 to 950 nm that included methane absorption bands. These images revealed the existence of a system of at least six stacked haze layers located above the upper cloud deck, southward of the hexagon. The presence of those hazes, and the knowledge of their physical properties, can provide important new clues to the understanding of the atmosphere of Saturn, and in particular, the hexagonal wave and its coupling with upper layers of the atmosphere.

In this paper, we concentrate first on the characterization of the hazes. We find that the vertical thickness of the individual layers ranged from 7 to 18 km, and that they extended in altitude above the pressure level of 0.5 bar up to 0.01 bar (~130 km in vertical extent), with a thin but extended aerosol layer above them, up to an altitude of ~340 km (pressure 0.4 mbar). Radiative transfer modeling of the spectral reflectivity of the hazes, including nadir-viewing images with the Hubble Space Telescope, shows that their properties can be explained by particles in the size range of 0.07–1.4 μm and number densities in the range of ~100–500 cm$^{-3}$. Then, we analyze the possible origin of the hazes. Their nature is compatible with their formation by condensation of hydrocarbon ices, including acetylene and benzene at the higher altitudes, and we propose that their vertical distribution could be due to upward-propagating gravity waves generated by dynamical forcing by the hexagon and its associated eastward jet.

## Results

**Multilayer stacked hazes above the hexagon.** Cassini ISS images of the limb on Saturn obtained with the Narrow Angle Camera (NAC)[14] on June 16, 2015 (northern spring season, solar longitude $Ls = 69°$) at high resolution (maximum of 1.25 km/pixel) showed a system of multilayered hazes above cloud tops (Fig. 1). The haze layers stand out conspicuously at 890 nm, in backward reflection at a strong methane absorption band. Additional images were obtained with a large number of filters at wavelengths ranging from ultraviolet (258 nm) to near infrared (938 nm)[14] (Supplementary Fig. 1, Supplementary Tables 1 and 2, and Methods). The limb images were taken in backward scattering under a Sun-atmosphere–spacecraft phase angle $\alpha = 29°$ (Supplementary Fig. 2). The planetographic latitude of limb in the images spans from 75.8°N to 77.1°N, just south of the eastward jet peak centered at 78.1°N that is embedded in the hexagon wave, thus on the anticyclonic side of the jet[4–7].

In Fig. 2, we show the vertical profile of reflectivity ($I/F$) of the haze layers at the different wavelengths observed. A small uncertainty of a few tens of km in the relative navigation of different images was corrected by taking into consideration the similarity of the main haze-layer structure in all filters. We assume the reference altitude at $z = 0$ km (where the optical depth at the line-of-sight $\tau_{LOS} \sim 1$) and a pressure level of $P = 0.5$ bar, based on a previous radiative transfer analysis of Cassini ISS images[15]. We used a vertical temperature profile at limb retrieved at latitude 77°N from Cassini Composite Infrared Spectrometer (CIRS) data (see Methods) in the same epoch (June 2015) and the hydrostatic condition to relate altitude ($z$) and pressure ($P$) and to calculate the layer thicknesses $\Delta z = H \ln\left(\frac{P_i}{P_{i+1}}\right)$ between two pressure levels ($P_i$, $P_{i+1}$). Here $H = \frac{R^*_g \langle T \rangle}{g} \approx 38 - 63$ km is the range of atmospheric-scale heights for the different altitudes covered, where $R^*_g = 3890$ J kg$^{-1}$ K$^{-1}$, $g = 9$ ms$^{-2}$, and $\langle T \rangle$ is the temperature at the altitude of the middle point between pressure levels $P_i$ and $P_{i+1}$[16].

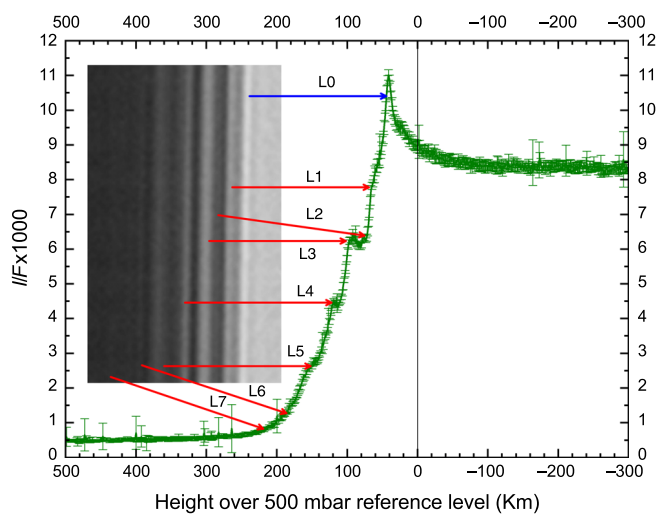

**Fig. 1 Haze layers in the hexagon area.** This graph shows the absolute reflectivity ($I/F$ times 1000) as a function of the vertical distance across the limb in the 890-nm methane absorption band filter. The inset shows an image of the haze layers in this filter and their identification (L0-L7) in the reflectivity scan across Saturn's limb. The haze layers extend meridionally at least from latitude +75.8°N to +77.1°N.

We have used both the photometric scans and direct measurements on contrast-enhanced versions of the images to determine the altitude of the boundaries for each individual haze layer, retrieving their mean altitudes (top and bottom) and vertical thicknesses (Table 1). The red and methane band images show the highest contrast and sharper separation between the layers. We identify at least seven haze layers above the limb, the first six layers with a thickness in the range of $\Delta z \sim 7$–18 km (~0.3 $H$). The uppermost layer extends above the 10-mbar pressure level and has a large vertical extent $\Delta z \sim 3$ $H$, suggesting a change in the atmospheric conditions at this altitude.

In order to survey the presence of hazes in other latitudes of the planet, we have performed a systematic search of images of the limb spanning the complete Cassini mission (December 2004–September 2017). We selected images with spatial resolution better than ~15 km pixel$^{-1}$, a limit we think reasonable to clearly capture hazes in reflected sunlight protruding at limb. We found images with these conditions obtained in a total of 120 days. These observations encompass 310 image series, each containing a sequence of images with a similar limb-viewing geometry but with different filters (1–15 filters, depending on the series). Only a limited part of these series contains all the filters used in our study (Supplementary Table 1). All latitudes were

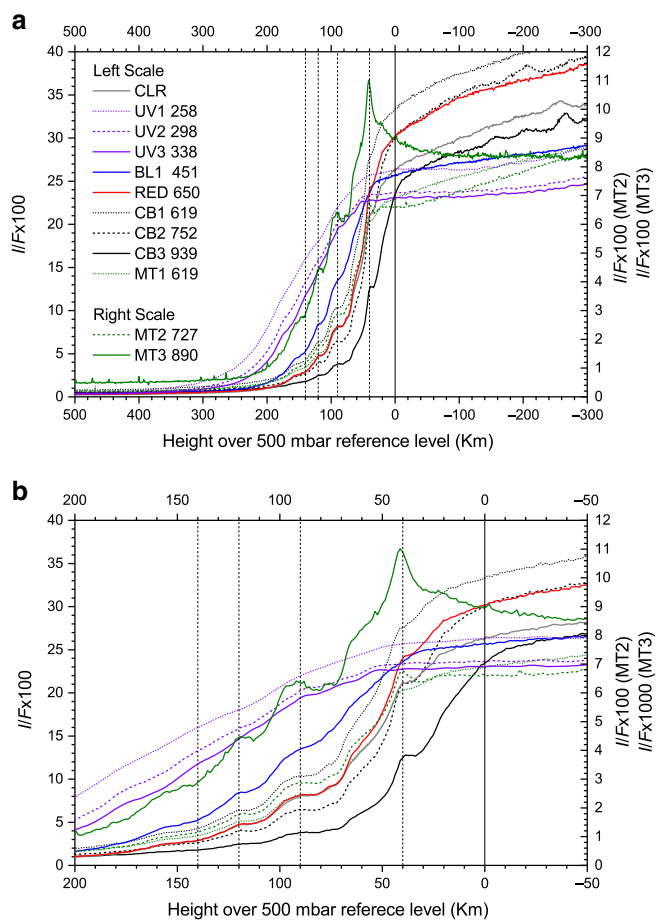

**Table 1 Location and thickness of the haze layers.**

| Layer | P (bar) bottom/top | Height (km) bottom/top | Δz (km) | H (km) bottom/top |
|---|---|---|---|---|
| L0 | 0.66/0.45 | −12.7/5 | 17.7 | 53/48 |
| L1 | 0.38/0.28 | 10.4/23.1 | 12.7 | 46/43 |
| L2 | 0.22/0.19 | 29.5/36.3 | 6.8 | 41/40 |
| L3 | 0.16/0.10 | 40.7/54.9 | 14.2 | 39/38 |
| L4 | 0.06/0.046 | 70.4/78.5 | 8.1 | 37/38 |
| L5 | 0.029 /0.025 | 99/105.2/111.4 | 12.4 | 39/41 |
| L6 | 0.017 /0.013 | 120/130 | 10 | 43/46 |
| L7 | 0.010 /0.0004 | 140/~340 | 200 | 48/64 |

*P and Height indicate the altitude location of the haze-layer edges relative to the reference level; Δz is the layer thickness, and H the atmospheric-scale height.*

**Fig. 2 Reflectivity scans of the haze layers at different wavelengths. a** Reflectivity profiles (I/F) extending from the limb to the upper atmosphere (sky background). **b** Zoom and detail of the reflectivity profiles close to the limb. Error bars in (I/F) for all filters are similar to those shown in Fig. 1 (not shown here for clarity). Note the change in the scale of the reflectivity according to the filters used (left and right y axis). Filters are identified by their Cassini ISS code (Supplementary Table 1) and their effective wavelength is given in nm. The vertical lines indicate the altitudes where the I/F values were used for the radiative transfer model.

observed at some time, but the temporal sampling, latitude coverage, spatial resolution, and viewing geometry were very variable (Supplementary Figure 3). The survey shows that haze layers are present above cloud tops at all latitudes and at any time, being their number, vertical extent, and reflectivity very variable. We have found only one clear case with a similar haze system formed by 5–6 stacked layers regularly separated. It was imaged on 24 November 2015 (6 months after the one studied here) and extends from latitudes ~66°N to 71.2°N (south of the hexagon region). It is likely that it forms part of the multilayer system studied here. Southward of ~65°N and up to latitude 41°N, the number of haze layers decreases to 3–4 and their thicknesses varies from one to another. The survey also shows that along the North Polar Region, i.e., from the hexagon to the pole, there are hazes formed by ~3–5 layers at the best available resolution of ~6.5–7 km/pixel. We cannot rule out the existence of thinner layers that cannot be resolved at the resolutions available at these latitudes. At other latitudes, the layer multiplicity and regular vertical distribution is not present. For example, from mid-latitudes to the equator, the number of layers reduces to 1–3 even at high resolution of 1–2 km pixel⁻¹. A general study of the haze layers we have observed at Saturn's limb along the whole Cassini mission will be presented elsewhere.

**Radiative transfer modeling.** The properties of the aerosol and haze layers were studied using two datasets. A first analysis was performed on multiwavelength images taken on 29–30 June 2015 with the Hubble Space Telescope (HST), 15 days after the Cassini limb images. The HST images[17] were obtained under a phase angle α = 3.9°, with a pixel scale of about 265 km and providing a nadir-integrated view of the region based on upper and upper cloud properties without resolving the limb. This nadir situation is analogous to that of our previous analysis of the region based on Cassini ISS images[15] obtained in 2013 under phase angles ranging from ~5.5° to 61.2°. An analysis of the HST nadir images explains the differences in brightness at different filters using a two-haze-layer model with (a) a thin layer in the stratosphere extending from ~5 to 200 mbar with an optical depth τ (0.9 μm) = 0.02 and particles with a mean size a = 0.15 μm; (b) a dense tropospheric haze extending from ~50 to 410 mbar with an optical depth τ (0.9 μm) = 12 and maximum particle number density $N_p$ ~ 35 cm⁻³ (Supplementary Figs. 4 and 5). These results agree with Sanz-Requena et al.[18], and may provide a better fit to 2013 observations than the models investigated in ref. [15], something that we are currently exploring and will be reported elsewhere. Because of the geometry of observations, the HST nadir analysis is unable to resolve each of the separate layers, and thus integrates into just two layers the properties of the entire multilayer system evident in the Cassini limb images.

A second analysis examines the wavelength-dependent reflectivity (I/F) of limb profiles obtained from Cassini (Fig. 2). We do this in two steps: first, estimating the particle density we expect for single scattering in the hazes, then, running a full multiple-scattering model. At wavelengths where the sensitivity to scattering and absorption by the gas is low, for optically thin layers with aerosol particles widely separated, the reflectivity is given by the single scattering of radiation[19,20]

$$\left(\frac{I}{F}\right)_{particles} \approx \bar{\omega}_0 \frac{p(\theta)}{4} \sigma_p N_p \sqrt{2\pi R_S \langle H \rangle} \qquad (1)$$

Here $R_S$ = 66,100 km is the Saturn's radius of curvature at the latitude of the hazes, which we approximated by the polar radius of curvature ($R_S \approx R_E^2/R_P$, being $R_E$ and $R_P$ the Equatorial and Polar radius, respectively), and we take a typical scale height for particles $\langle H \rangle$ = 15–20 km (Table 1). From a previous study of Saturn's stratospheric particles in this region[15], we take a single-scattering albedo $\bar{\omega}_0$ = 0.9 and a scattering phase function $p(\theta)$ = 0.05 at scattering angle $\theta$ = 150°[15]. The scattering cross section is $\sigma_p = Q_s \pi a^2$, with the scattering efficiency factor $Q_s$ ~ 0.1, and the particle radius is taken to be a = 0.15 μm from HST analysis. At $z = 100$ km, Fig. 2 shows that $(I/F) \sim 0.05$ at the wavelength of the continuum red filters sensitive to the particles, which yields a number density for the particles of the order of $N_p \sim 220$ cm$^{-3}$.

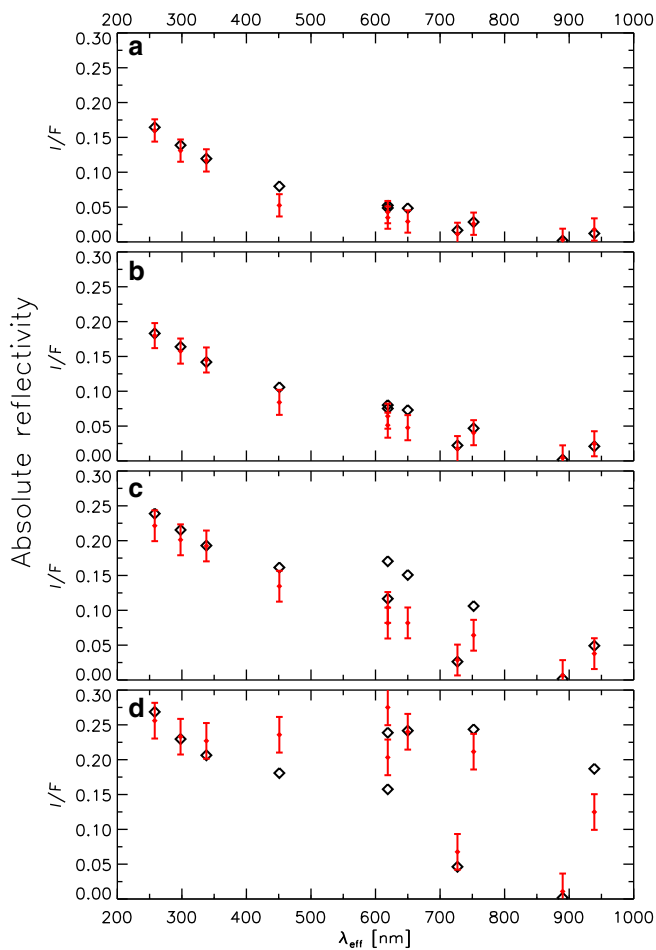

**Fig. 3 Radiative transfer model fit to the limb reflectivity scans.** Panels show the reflectivity spectra from Fig. 2 (rhombs) and the limb-model results for the hazes at different altitudes (red dots with error bars). **a** $z = 140$ km, $N_p = 500$ cm$^{-3}$, $H_{aerosol} = 19$ km. **b** $z = 120$ km, $N_p = 250$ cm$^{-3}$, $H_{aerosol} = 20$ km. **c** $z = 90$ km, $N = 100$ cm$^{-3}$, $H_{aerosol} = 21$ km. **d** $z = 40$ km, $N_p = 100$ cm$^{-3}$, $H_{aerosol} = 18$ km.

The complete analysis of the properties of each layer observed at the limb was aimed to reproduce the measured spectral reflectivity, and uses a model that solves the radiative transfer problem under spherical geometry[21]. We selected four tangential altitudes for the spacecraft's line of sight, namely $z = 40$, 90, 120, and 140 km. The first one probes the lower tropospheric haze (Fig. 2), and the other three probe the stratospheric hazes. Our model considers an exponential drop in the particles' number density from the troposphere and stratosphere, with different optical properties at each of the two regions. We considered as free parameters the number density of particles at the troposphere ($N_p$) and the corresponding scale height ($H_{aerosol}$). In the stratosphere, we used a wavelength-dependent single-scattering albedo based on that for Titan's hazes[22] because the chemical composition of both hazes is consistent with hydrocarbons (see the next section and Methods). In Fig. 3 we show the best fits of the limb models to the reflectivity profiles. For the tropospheric haze at $z = 40$ km (Fig. 3d), we find $H_{aerosol} = 18$ km, $N_p = 100$ cm$^{-3}$, and $a = 1.45$ μm ($Q_s \sim 1$). For the stratospheric hazes, we find $H_{aerosol} = 19–21$ km, $N_p = 100–500$ cm$^{-3}$, and $a = 0.07$ μm ($Q_s \sim 0.1$) (see the Methods section). These particle number densities are consistent with the estimates using the single-scattering approach given above.

**Nature of the hazes.** Given that the haze boundaries are sharp (Fig. 1), a likely scenario is that the haze particles form by a condensation process at the level where saturation of a given compound occurs. At the cold temperatures of the lower stratosphere of Saturn, hydrocarbons tend to condense out as ices[10,23,24]. High in the atmosphere ($z > 300–350$ km), the haze particles could be a mix of solid organics formed by methane photolysis and electron deposition[25] or, alternatively, from a combination of auroral activity and ionospheric ion chemistry[26]. Saturn's main auroral oval, associated with the boundary between open and closed magnetic field lines, occurs at an average latitude of 76.5°N[27,28], corresponding to the location of the hazes studied here. Cassini CIRS observations showed haze signatures at only two locations in the polar area (latitudes 80°S[29] and 77°N[30]), i.e., in the auroral oval region. One must consider therefore the possibility that the hazes above the hexagon are due to the energy deposition by the aurora[8]. However, the detection of multilayer hazes outside the aurora region, described above, points to different mechanisms. Perhaps, the highest ($z > 140$ km) and most extended vertical layer (L7 in Table 1) could have an auroral origin. We do not disregard that the hazes can also be formed by photochemical tholins and fractal aggregate particles as it has been proposed for similar hazes in Titan[31,32]. Since the hazes we report lay in the upper troposphere and lower stratosphere ($z < 340$ km), we explore their possible formation by ice condensation of photochemically produced hydrocarbons[33].

We used vertical profiles of the temperature and hydrocarbons' volume-mixing ratio retrieved from Cassini/CIRS thermal emission spectra obtained at the same epoch (June 16, 2015) and latitude (77°N) as in the haze images (see Methods). In Fig. 4 we show a set of hydrocarbon candidates with their vapor-pressure curves[24] and condensation altitudes (Supplementary Fig. 6). The retrieved mole fractions for each compound are consistent with expectations from a photochemical model, except for benzene, which is underestimated in current photochemical models, at least at polar latitudes[29]. We find that $C_3H_8$ (propane), $C_2H_2$ (acetylene), $CH_3C_2H$ (propyne), $C_4H_2$ (diacetylene), and $C_6H_6$ (benzene) are good candidates to form hazes from condensation at different stratospheric levels. This agrees with previous studies that also considered $C_5H_{12}$ (pentane) and $C_4H_{10}$ (butane) as condensates in the 5–50-mbar level[25]. It has been suggested that because of the proximity of the condensation altitude levels of these compounds and due to the gravitational settling of the particles, they could condense on one another forming layers of mixed composition[25]. Condensation of water ice could also occur at high altitudes in a thin layer above 0.1-mbar pressure level[25]. At deeper levels, in the upper troposphere, $NH_3$ ice is a good candidate to form the lower haze[34].

**Vertical propagation of gravity waves.** The quasi-regular vertical distribution of the multiple haze layers suggests that they could be produced by the passage of vertically propagating waves, inducing vertical oscillations in the temperature profile, favoring the condensation of compounds at specific altitudes according to the saturation vapor-pressure curves of the compounds.

As indicated above, the haze layers we study here were observed between planetographic latitudes 75.8°N and 77.1°N, but these limits are imposed by the limited field of view of the limb by Cassini ISS at that time. The hazes most probably extended beyond these latitudes at least south of the hexagon wave and its embedded jet down to latitude 66°N, as shown by our survey of Cassini ISS images during the Cassini mission

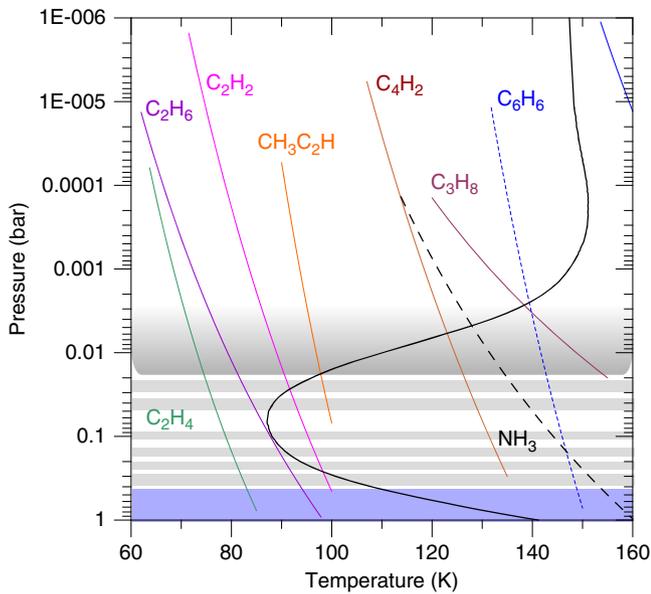

**Fig. 4 Vertical temperature profile and saturation vapor-pressure curves for different hydrocarbons.** The reference altitudes used for the assumed hydrocarbon mole fractions are given as follows (Supplementary Fig. 6): $C_3H_8$ ($10^{-7}$ at $P = 3$ mbar); $C_4H_2$ ($3.8 \times 10^{-12}$ at $P = 3$ mbar); $CH_3C_2H$ ($7 \times 10^{-11}$ at $P = 3$ mbar); $C_2H_2$ ($4.8 \times 10^{-8}$ at $P = 3$ mbar); $C_2H_6$ ($4.5 \times 10^{-6}$ at $P = 3$ mbar); $C_2H_4$ ($3 \times 10^{-10}$ at $P = 3$ mbar). For $C_6H_6$, two values of the mole fraction are considered at two altitudes, with the corresponding saturation curves represented by two lines: blue continuous (mole fraction $7.7 \times 10^{-12}$ at $P = 3$ mbar, from ref. [33]) and blue dashed (mole fraction $2.6 \times 10^{-10}$ at $P = 0.1$ mbar from Supplementary Fig. 6). Condensation and haze base formation occur at altitudes where the saturation vapor-pressure curves intersect with the temperature profile. The location of the observed haze layers is indicated by horizontal gray bands.

described above. Here we focus on the latitudes on which the multilayer system has been observed in more detail as shown in Table 1. At cloud level ($\sim$0.5–1 bar), the eastward jet shows velocities increasing from $u \sim 5$ ms$^{-1}$ at 75.8°N to $u \sim 55$ ms$^{-1}$ at 77.1°N, in the anticyclonic side of the jet[5–8]. The hexagon and its meandering eastward jet have been proposed to be a trapped Rossby wave[5,35]. We explore the possibility that the hexagon and jet force gravity waves at the cloud level that, propagating vertically in the region of positive static stability[12,36], generate the conditions for the stacked haze-layer formation. There are abundant descriptions in the literature of non-orographic gravity wave generation on Earth from jets and fronts[37] that could be applicable to the Saturn hexagon and, at other latitudes, by Saturn's zonal jet system. The vertical propagation of hydrostatic gravity waves requires that the angular frequency $\omega$ of the wave verifies

$$|f| \leq |\omega| \leq |N| \qquad (2)$$

where $f = 2\Omega \sin \varphi = 3.17 \times 10^{-4}$ s$^{-1}$ is the Coriolis parameter at latitude $\varphi = 76.5°$N and $N \approx 0.012$ s$^{-1}$ the Brunt–Väisälä frequency[13]. We consider linear perturbations to the governing dynamic equations[20,38–40], and search for plane-wave solutions of the form $\psi(x, y, z, t) = \psi_0 \exp[(z/2H) + i(kx + \ell y + mz - \omega t)]$, where $\omega = (c_x - u)k$ is the intrinsic wave frequency, $m = 2\pi/L_z$ denotes the vertical wavenumber, with $L_z$ being the vertical wavelength, and the horizontal wavenumbers are $k = 2\pi/L_x$, $\ell = 2\pi/L_y$, with $L_x$ and $L_y$ being the horizontal

wavelengths. The waves obey the dispersion relationship[20,38–40]:

$$m^2 = \frac{N^2 - \omega^2}{\omega^2 - f^2} k^2 \delta^2 - \frac{1}{4H^2} + \left( \frac{\omega^2 - \omega_a^2}{c_s^2} \right) + \frac{k}{\omega} \left( \frac{1}{H} \frac{du}{dz} + \frac{d^2u}{dz^2} \right) \qquad (3)$$

where $\delta^2 = \left[ 1 + (\ell/k)^2 \right]$. Using $H = 40$ km as the average gas-scale height, the speed of sound is $c_s = \sqrt{\gamma R^* T} = 710$ ms$^{-1}$ ($\gamma = 1.4$, $R^* = 3890$ J kg$^{-1}$ K$^{-1}$, $T = 90$ K) and the acoustic frequency $\omega_a = c_s/2H \approx 0.01$ s$^{-1}$. We assume that the atmosphere is homogeneous since at the latitudes and epoch of the observations, the dependence of $N$ and $u$ with altitude is small above $\sim$0.2 bar (Fig. 10b, c in ref. [13]), thus dropping the last term in Eq. (3).

The wave frequency $\omega$ is calculated using $c_x \approx 0$ ms$^{-1}$ (hexagon-phase speed) and $u = 120$ ms$^{-1}$ (value of the peak of the eastward jet)[5,6]. We take for the horizontal scale the Rossby deformation radius $L_D \approx 1500$ km because of the prevalence of geostrophic conditions at the hexagon[5,6,36]. Then $k = \ell \approx 4 \times 10^{-6}$ m$^{-1}$, $\delta = 1.41$, and $\omega = (c_x - u)k \approx 5 \times 10^{-4}$ s$^{-1}$, implying that $\omega^2 \ll \omega_a^2$. The acoustic term and the vertical damping scale have similar values in Eq. (3), i.e., $\left| (\omega^2 - \omega_a^2)/c_s^2 \right| \sim 1/4H^2 \approx 2 \times 10^{-10}$ m$^{-2}$ and therefore the first term dominates in Eq. (3). Then we have

$$m^2 \approx \frac{N^2 - \omega^2}{\omega^2 - f^2} k^2 \delta^2 \approx 3 \times 10^{-8} \text{ m}^{-2} \qquad (4)$$

Figure 5 (upper panel) shows the dependence of the vertical wavelength $L_z = 2\pi/m$ for gravity waves as a function of the frequency and for two values of the intrinsic velocity $(c_x - u)$ according to Eq. (4). The frequency range is bounded by the Coriolis and Brunt–Väisälä frequencies from Eq. (2). The expected values for the vertical wavelength would be in the range of $\sim$20–60 km. Figure 5 (bottom panel) shows an idealization of the vertical structure of the amplitude of propagating internal waves with no dissipation (i.e., without damping by eddy and molecular viscosity). The amplitude grows upward $\sim\exp(z/2H)$ to make energy density constant with altitude[38]. For hazes produced by condensation, it can be expected that the vertical wavelength is $L_z \sim 2\Delta z$ with $\Delta z$ the layer thickness. Taking $\Delta z \sim 8$–18 km (the range of thicknesses for layers L0–L6 in Table 1) gives $L_z \sim 16$–36 km, consistent with the above calculations. The maximum and minimum of the amplitude oscillation are partially in agreement with the observed pattern in some of the layers (L0–L2, L4, and L6), but differ from these idealized oscillations in others (L3 and L5). Obviously, this is to be expected, since in our analysis we neglect dissipation effects and changes in the scale height. In addition, compression and rarefaction due to the gravity wave passage could affect the particle number density and thickness of the haze layers[19,20]. However, the lack of data on both the mechanisms subjacent to the gravity wave generation and the initial upward velocity of the parcels in the wave, does not allow us to deepen the analysis.

## Discussion

We have analyzed in detail a system of at least seven detached haze layers in the upper troposphere and stratosphere above the hexagon's clouds using Cassini ISS high-resolution limb images. The nature of these hazes is compatible with their formation by condensation of a number of hydrocarbons detected in Saturn's atmosphere forming high in the atmosphere by photochemical reactions (e.g., $C_2H_2$, $C_4H_2$, $C_6H_6$, and $CH_3C_2H$). We cannot rule out the possibility that the hazes are formed by photochemical tholins and fractal aggregate particles from more complicated aerosol microphysics and chemistry as it has been proposed for similar hazes in Titan[31,32]. We suggest that the vertical distribution of the stacked system of haze layers could be produced by the upward propagation of internal gravity waves excited in the hexagon wave and its eastward jet area, perhaps in a similar way to what it occurs in Earth's jets and fronts[37]. In fact, the lower value of the derived vertical wavelength, in the range of 20–60 km, is compatible with the retrieved aerosol-scale height and with the measured layer thicknesses if condensation is the mechanism behind haze formation.

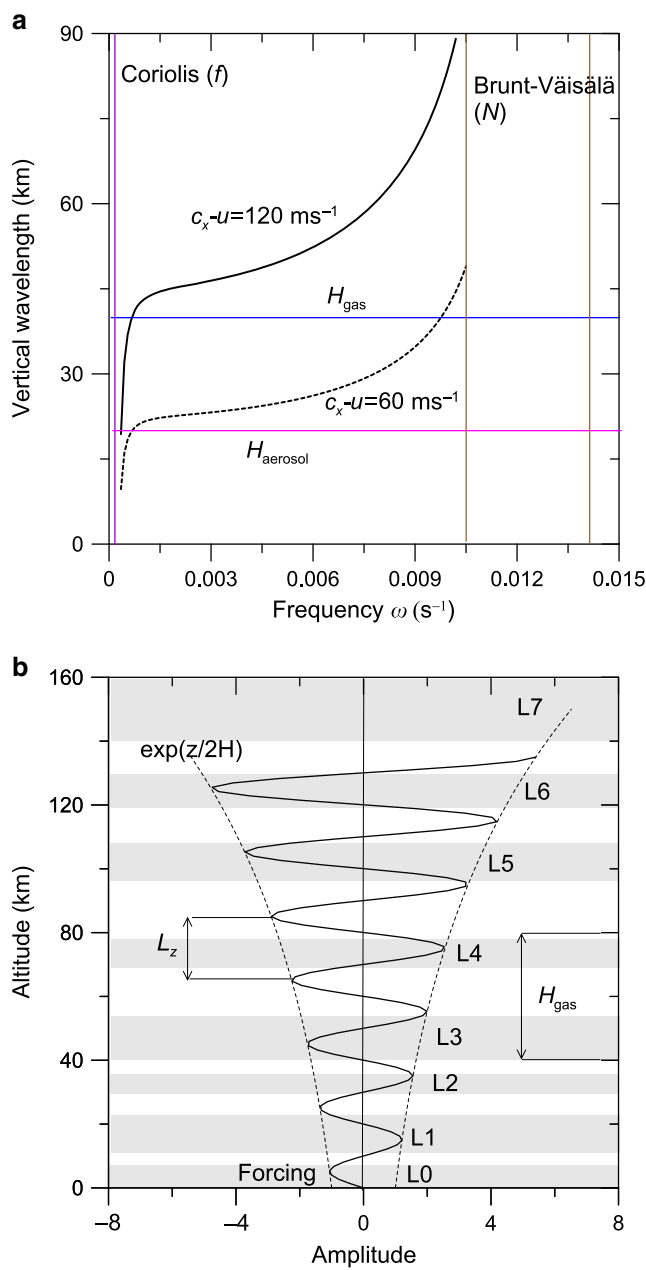

**Fig. 5 Gravity waves and haze layers. a** Vertical wavelength as a function of frequency. The frequency range is between Coriolis $f$ (violet line) and Brunt-Väisälä frequency $N = 0.012$ s$^{-1}$ (brown line) with a range of measured values $N = 0.01$-$0.015$ s$^{-1}$ taken from ref. [13]. Two cases are shown for values of the intrinsic velocity given by ($c_x - u$) = 60 ms$^{-1}$ (dashed line) and 120 ms$^{-1}$ (continuous line) [5,6,13]. The mean-scale height of the atmosphere ($\langle H_{gas}\rangle = 40$ km) and the average aerosol-scale height ($\langle H_{aerosol}\rangle = 20$ km) as derived from the radiative transfer model calculations are also indicated. **b** Idealized amplitude oscillations for the gravity wave model assuming a mean vertical wavelength $L_z = 20$ km and $H_{gas} = 40$ km, and no dissipation. The altitude location and thickness of the observed haze layers are also indicated for comparison as given in Table 1.

Our survey of Saturn's limb using Cassini ISS images along the whole mission (2004–2017) shows that haze layers exist in all latitudes above the upper cloud deck, but multilayer structures are only found at high northern latitudes, and the set described here has the best resolution and filter coverage. Ongoing studies will characterize and determine the properties of haze layers in Saturn according to their latitude and in relation to zonal jets. The generation of gravity waves by these jets will be explored by taking into account jet wind intensity and meridional distribution[41], as well as their variations in altitude and over time in some latitudes[42]. Finally, the future research should also explore microphysical models, the growth, and destruction mechanisms of the observed aerosols, and look for changes in the haze properties in relation to the seasonal insolation cycle[10].

## Methods

**Cassini images.** The Cassini ISS images here analyzed are public data retrieved from the Atmospheres node of the Planetary Data System (PDS) NASA. Images are available from the following filters: UV1, UV2, UV3, BL1, RED, MT1, MT2, MT3, CB1, CB2, and CB3 (see details in ref. [14] and Supplementary Table 1). The maximum resolution achieved was in the range of 2–6 km/pixel, but images with ultraviolet filters were binned (UV1 ×4, UV2–UV3 ×2). During the observations, the phase angle varied from 29.1° to 30.5° (corresponding to a scattering angle $\theta \sim$ 150°, see Supplementary Fig. 2). The images were calibrated with the CISSCAL software[43] and navigated to determine the limb location and obtain reflectivity scans across the limb using the software PLIA[44].

**HST images.** Images used in this work are the same as in ref. [17], where all relevant information is detailed. The following filters were available: F225W, F336W, F410M, F502N, F547M, F689M, FQ727N, FQ750N, FQ889N, and FQ937N (details given in Supplementary Table 1). All HST images were calibrated in absolute reflectivity as explained in ref. [45]. Image navigation and reflectivity scans were obtained using the software LAIA[46]. Reflectivity was evaluated at 35 points at the latitude of interest from −80° to +80° from the central meridian, providing coverage of all the available illumination and viewing geometries within a plane-parallel approximation.

**Radiative transfer.** The radiative transfer analysis used two different codes: one for the spherical geometry case[21] and the other, the NEMESIS code, for nadir case[47]. For the interpretation of the limb-viewing observations, we used a Monte Carlo model[21] that solves the multiple-scattering problem in a spherical-shell atmosphere. The setup of the model is similar to that described in a previous work[48], and takes as input the geometry of the observations and the optical properties of the atmosphere. We adopted a background gas atmosphere specified by the profiles of the Rayleigh scattering coefficient and methane absorption, and based on ref. [15]. On top of the gas optical properties, we specified an exponential profile for the number density of the particles. The profile is described as $N_p(z) = N_p \exp(-z/H_{aerosol})$, where $N_p$ and $H_{aerosol}$ are considered to be tunable parameters, and $z$ is measured from the tropopause.

The radiative transfer solution depends on the scattering-phase function, the single-scattering albedo, and the extinction cross section of the particles. We have conducted a sensitivity analysis with the spectral distribution measured at $z = 140$ km to demonstrate the impact of some of the corresponding choices.

- Perturbation of the extinction cross section. Because in the radiative transfer problem the extinction cross section of the particles always appears multiplied by the number density of the particles, both factors are inversely related, which affects directly our inferred $N_p$. Smaller cross sections will result in larger $N_p$ values, and vice versa.

- Perturbation of the scattering-phase function. Alternatively, to our standard choice for the scattering-phase function, we have implemented the scattering-phase function corresponding to Titan's stratospheric haze at a wavelength of 350 nm. For $N_p = 500$ cm$^{-3}$ and $H_{aerosol} = 19$ km, a good fit is still obtained, but the model appears too indifferent at the shorter wavelengths.

- Perturbation of the single-scattering albedo. We have modified our standard choice for the single-scattering albedo by scaling it by a 0.9× factor at all wavelengths. For $N_p = 500$ cm$^{-3}$ and $H_{aerosol} = 19$ km, the fit actually improves, which suggests that Saturn's stratospheric haze may be darker than that in our models.

In the troposphere ($z < 0$), we assumed the same scattering cross sections as for the nadir analysis. In the stratosphere ($z > 0$), we parameterized the cross section as $\sigma_p(\lambda) = \sigma_{Titan}(939 \text{ nm}) \cdot (939 \text{ nm}/\lambda(\text{nm}))^4$, with $\sigma_{Titan}(939 \text{ nm}) = 1.28 \times 10^{-11}$ cm$^{-2}$ [23]. For the scattering properties, we adopted for all wavelengths the scattering-phase function proposed for Titan's stratospheric particles at 650 nm[22], which is in the mid of the range of wavelengths explored here. We kept the single-scattering albedo of the particles as an additional degree of freedom for the final fine-tuning of the model-observation comparison. The models presented in the paper adopt for the stratospheric particles a single-scattering albedo that is 0.5 for $\lambda \leq 450$ nm, and 0.9× the single-scattering albedos proposed for Titan's stratospheric particles for $\lambda > 450$ nm[22]. Other choices in how the haze optical properties are described in the radiative transfer model will have an impact on the inferred number densities. We consider our estimated values to be valid first-order descriptions that could eventually be refined by, for instance, combining observations at other phase angles.

The NEMESIS code uses an optimal estimator scheme to find the most likely values of atmospheric parameters starting from a priori assumption. Our a priori assumption for this work was inherited from previous studies at neighboring latitudes[49], where all the details in the initialization of the model atmosphere can be found. Further details on the goodness of fit and retrieved parameters from the nadir observations can be found in the companion Supplementary Table 2 and Supplementary Figs. 3 and 4.

**Temperature and hydrocarbon vertical profiles.** Vertical profiles of the temperature and volume-mixing ratio of six hydrocarbons were obtained from the analysis of thermal emission spectra recorded by Cassini/CIRS using a radiative transfer model coupled to a retrieval algorithm. CIRS was a Fourier-transform spectrometer onboard the Cassini Spacecraft, covering the spectral range of $600-1400$ cm$^{-1}$ (7–17 μm). We focus here on spectra recorded in limb-viewing geometry at planetographic latitude 77°N on 16 June, 2015, at a spectral resolution of 1.5 cm$^{-1}$. We first exploited emission features from the v4 band of CH$_4$ (centered at 1304 cm$^{-1}$) and of the emission induced by collisions by H$_2$–H$_2$ and H$_2$–He (range 600–660 cm$^{-1}$) to constrain the temperature vertical profile, following the methodology described in ref.[50]. As thermal emission depends on both the temperature and the abundance of these molecules, we assume a methane vertical profile taken from a photochemical model and scaled to match a deep mixing ratio of $4.5 \times 10^{-3}$[51] and a He mixing ratio of 0.1355[52]. By combining typically 10 spectra acquired at different tangent altitudes, we constrain the temperature between ~20 mbar and 3 μbar with a vertical resolution of one atmospheric-scale height (~50 km). Once the temperature is set, we then analyzed emission bands from ethane (centered at 822 cm$^{-1}$), acetylene (730 cm$^{-1}$), propane (748 cm$^{-1}$), diacetylene (628 cm$^{-1}$), methyl acetylene (634 cm$^{-1}$), and benzene (673 cm$^{-1}$) to derive their volume-mixing ratio vertical profile[29,51,53]. Vertical sensitivity extends from 3 mbar to 3 μbar, except for propane (only constrained between 3 and 0.3 mbar) and benzene (retrieved between 0.1 mbar and 3 μbar).

## Data availability
Cassini ISS images are available via NASA's Planetary Data System (PDS): http://pds-atmospheres.nmsu.edu/data_and_services/atmospheres_data/Cassini/iss.html. The data that support the analysis and plots within this paper and other findings of this study are available from the corresponding author upon request.

## Code availability

## Acknowledgements
This work has been supported by the Spanish project AYA2015-65041-P (MINECO/FEDER, UE) and Grupos Gobierno Vasco IT1366-19. We acknowledge the Cassini ISS team for obtaining the limb images. This work used data acquired from the NASA/ESA HST Space Telescope, which is operated by the Association of Universities for Research in Astronomy, Inc., under NASA contract NAS 5-26555. We acknowledge the three orbits assigned by the Director Discretionary time from HST for this research (GO/DD Program 14064, PI A. Sánchez-Lavega).

## Author contributions
A.S.L. directed the work, measured haze altitudes, studied haze nature, and the dynamical interpretation. T.d.R. made the limb photometry and together with J.P. discussed the dynamical model. The radiative transfer analysis was performed by A.G.M. on Cassini limb images, and by J.F.S.R. and S.P.H. on HST nadir images, with ASL contribution. S.G. performed the temperature and hydrocarbon retrievals. R.H. developed PLIA software. All authors discussed the results and contributed to preparing the paper.

## Competing interests
The authors declare no competing interests.

## Additional information
**Supplementary information** is available for this paper at https://doi.org/10.1038/s41467-020-16110-1.

**Correspondence** and requests for materials should be addressed to A.S.-L.

**Peer review information** *Nature Communications* thanks Michael Roman, Robert West and the other, anonymous, reviewer(s) for their contribution to the peer review of this work.

**Reprints and permission information** is available at http://www.nature.com/reprints

**Publisher's note** Springer Nature remains neutral with regard to jurisdictional claims in published maps and institutional affiliations.